\newcommand{\diracslash}[1]{#1\llap{/\kern2pt}}

\newcommand{\be}{\begin{equation}}
\newcommand{\ee}{\end{equation}}
\newcommand{\bea}{\begin{eqnarray}}
\newcommand{\eea}{\end{eqnarray}}
\newcommand{\ba}[1]{\begin{array}{#1}}
\newcommand{\ea}{\end{array}}

\documentclass[prd,aps,floats,nofootinbib,tightenlines,showpacs]{revtex4}
\usepackage{epsfig,graphicx,pstricks}
\usepackage{psfrag}
\usepackage{color}
\usepackage{amsmath}
\usepackage{amsfonts}
\usepackage{amssymb}
\usepackage{textcomp}
\usepackage{multirow}
\usepackage{subfigure}
\usepackage{slashed}

\begin{document}

\title { Thermodynamic properties of  gluon plasma: A q-potential approach}
\author{Guruprasad Kadam}
\affiliation{Department of Physics and Material Science and Engineering,  Jaypee Institute of Information Technology, A-10, Sector-62, Noida, UP-201307, India}
\email{guruprasadkadam18@gmail.com}

\date{\today} 

\def\be{\begin{equation}}
\def\ee{\end{equation}}
\def\bearr{\begin{eqnarray}}
\def\eearr{\end{eqnarray}}
\def\zbf#1{{\bf {#1}}}
\def\bfm#1{\mbox{\boldmath $#1$}}
\def\hf{\frac{1}{2}}
\def\sl{\hspace{-0.15cm}/}
\def\omit#1{_{\!\rlap{$\scriptscriptstyle \backslash$}
{\scriptscriptstyle #1}}}
\def\vec#1{\mathchoice
        {\mbox{\boldmath $#1$}}
        {\mbox{\boldmath $#1$}}
        {\mbox{\boldmath $\scriptstyle #1$}}
        {\mbox{\boldmath $\scriptscriptstyle #1$}}
}

\begin{abstract}
In this work, we study the thermodynamic properties of quark-gluon plasma using Kramer's $q-potential$ method. We propose modification in the first law of thermodynamics by including  temperature dependent term when single particle energies themselves are temperature dependent.  With this modified first law, we derive an expression for pressure starting from Kramer's  $q-potential$. We find that, to maintain thermodynamic consistency,  pressure receives an additional term solely due to medium dependent dispersion relation. The energy density, by definition, remains a sum over all the single particle energies with appropriate weight factor.  We confront this quasiparticle model with the  lattice QCD data of SU(3) pure gluon plasma. 
\end{abstract}


\maketitle

\section{Introduction}
Quasiparticle model is a phenomenological model  used to describe the thermodynamics of interacting  quark-gluon-plasma (QGP)  In this model the interacting gas of massless quarks and gluons can be effectively described by ideal gas of massive quarks and gluons\cite{Goloviznin:1992ws}. This behavior is the upshot of refractive nature of the medium so that all the excitations change their dispersion relation. Effective  mass of quarks and gluons  has a dynamical origin and it depends on thermodynamic parameters, $viz.$, temperature (T) and chemical potential ($\mu$) as well as momentum ($\bf{k}$). 

Initial quasiparticle models\cite{Goloviznin:1992ws, Peshier:1994zf} failed to accurately explain the lattice QCD data. Gorenstein and Yang\cite{Rischke:1992uv, Rischke:1992rk} later pointed out that these models are thermodynamically inconsistent and proposed a remedy by introducing a temperature dependent vacuum energy.  Ref.\cite{Bannur:2006hp}  showed that this  exercise is unnecessary and following the standard statistical mechanics (SM),  proposed a new quasi-particle model. This single parameter model is also thermodynamically consistent and explains lattice data very well. In this work, we propose an alternative, thermodynamically consistent, quasiparticle model.  We achieve this by modifying the first law of thermodynamics. We include temperature dependent term  in the first law  when single particle energies themselves are temperature dependent.  With this modified first law, we derive an expression for pressure starting from Kramer's  $q-potential$. We shall show that the pressure receives additional terms such that the thermodynamic consistency is maintained without changing the definition of energy density.

 We organize the paper as follows. In Sec. \ref{secII}  we describe our quasiparticle model and also show its thermodynamic consistency.  In Sec. \ref{secIII} we estimate the thermodynamic properties of quark gluon plasma within the ambit of proposed quasiparticle model and compare it with the lattice gauge theory simulations of pure SU(3) gluodynamics. Finally in Sec. \ref{secIV} we summarize and conclude.

\section{Quasiparticle model}
\label{secII}
Thermodynamic system in which energy ($E_{r}$) as well as number of particles $(N_{s})$ of the subsystems fluctuate about mean values but temperature (T) and chemical potential ($\mu$) are fixed can be described by the grand canonical ensemble (GCE).  The fundamental quantity from which all the thermodynamic quantities can be derived is the grand canonical partition function $\mathcal{Z}$ given by\cite{pathria2017statistical}

\be
\mathcal{Z}=\sum_{s,r}e^{-\beta E_{r}-\alpha N_{s}}
\ee

where $\beta=1/T$ and $\alpha=-\beta\mu$. 
 
 In statistical mechanics, only $\bar E$ and $\bar N$ are defined; all other quantities are derived.  $\bar E$ and $\bar N$ are defined as
 
 \begin{equation}
 	\bar E=\frac{1}{\mathcal{Z}}\sum_{s,r}E_s e^{-\beta E_{r}-\alpha N_{s}}
 	\label{ebar}
 	\end{equation}
 
  \begin{equation}
 	\bar N=\frac{1}{\mathcal{Z}}\sum_{s,r}N_r e^{-\beta E_{r}-\alpha N_{s}},
 	\label{nbar}
 \end{equation}
 respectively.  The connection between the statistics of the grand canonical ensemble and the thermodynamics of the system can be established using   Kramer's $q-potential$  defined as\cite{pathria2017statistical}
 
 \begin{equation}
 	q\equiv \text{ln} \bigg[\sum_{r,s}\text{exp}(-\alpha N_r-\beta E_s) \bigg]
 	\label{qdef1}
 	\end{equation}
 
 Quantity $q$ is a function of the parameters $\alpha$, $\beta$ and also of all the $E_s$. The total differential of $q$ is written as

 \begin{equation}
 	dq=\bigg(\frac{\partial q}{\partial \alpha}\bigg)_{\beta,\{E_s\}}d\alpha+\bigg(\frac{\partial q}{\partial \beta}\bigg)_{\alpha,\{E_s\}}d\beta+\sum_{r,s}\bigg(\frac{\partial q}{\partial E_s}\bigg)_{\alpha,\beta}dE_s
 \end{equation}

 Using (\ref{ebar}) and (\ref{nbar}) we get

 \begin{equation}
 	dq=-\bar Nd\alpha-\bar E d\beta-\beta \bigg<\frac{\partial E_s}{\partial \beta}\bigg>d\beta-\frac{1}{\mathcal{N}}\sum_{r,s}<n_{r,s}>dE_s
 	\label{dq1}
 \end{equation}

 where the angled bracket corresponds to statistical average of the quantity inside\footnote{We shall use overhead bar and the angled bracket interchangeably to denote statistical averages(see Eqs. (\ref{ebar}) and (\ref{nbar})).}. $n_{r,s}$ corresponds to number of members in an ensemble of $\mathcal{N}$ members having energy $E_s$ and number of particles $N_r$. We can rewrite Eq.(\ref{dq1}) as
 
 \begin{equation}
 	d(q+\alpha\bar N+\beta \bar E)=\beta(d\bar E+dW_0-\mu d \bar N)-\beta \bigg<\frac{\partial E_s}{\partial \beta}\bigg>d\beta
 	\label{dq2}
 	\end{equation}
 
 where we have made following correspondence:
 
 \begin{equation}
 	\delta W_0=-\frac{1}{\mathcal{N}}\sum_{r,s}<n_{r,s}>dE_s, \hspace{0.5 cm} \mu=-\frac{\alpha}{\beta}
 	\end{equation}
 	
 	In the absence of quasiparticles, the first law of thermodynamics is written as
 	
 	\begin{equation}
 		d\bar E=T\:dS-dW_0+\mu d\bar N
 		\label{1stlaw0}
 	\end{equation}
 	
 	However, if the single-particle energies depend on temperature, then there must be a term in the first law of thermodynamics which accounts for the change in  internal energy due to medium-dependent quasiparticle energies. In our case, the first law of thermodynamics can be rewritten as
 	
 	\begin{equation}
 		d\bar E=T\:dS-dW_0-\frac{1}{\beta}d\mathcal{Q}_\beta+\mu d\bar N
 		\label{1stlaw1}
 	\end{equation}
 	where $d\mathcal{Q}_\beta$ depends on temperature, the exact form of which can be obtained using thermodynamic consistency conditions (see Eq. (\ref{Qbeta})). Using (\ref{1stlaw1})  in Eq.(\ref{dq2}) we get

  \begin{equation}
 	d(q+\alpha\bar N+\beta \bar E)=\beta(T dS-\frac{1}{\beta}d \mathcal{Q}_\beta)-\beta \bigg<\frac{\partial E_s}{\partial \beta}\bigg>d\beta
 \end{equation}

Integrating, we get

 \begin{equation}
 	q+\alpha\bar N+\beta \bar E=S- \mathcal{Q}_\beta-\int \beta \bigg<\frac{\partial E_s}{\partial \beta}\bigg>d\beta
 	\end{equation}
 
 or

  \begin{equation}
 	q=\beta(TS+\mu\bar N-\bar E-\frac{1}{\beta}\mathcal{Q}_\beta)-\int \beta \bigg<\frac{\partial E_s}{\partial \beta}\bigg>d\beta
 \end{equation}
 
 Using thermodynamic relation, $\bar E=TS-PV-\frac{1}{\beta}\mathcal{Q}_\beta+\mu\bar N$,  we get
 
 \begin{equation}
 	\frac{PV}{T}=q+\int \beta \bigg<\frac{\partial E_s}{\partial \beta}\bigg>d\beta
 	\end{equation}
 
 or 
 
  \begin{equation}
 	\frac{PV}{T}=\text{ln} \bigg[\sum_{r,s}\text{exp}(-\alpha N_r-\beta E_s) \bigg]+\int \beta \bigg<\frac{\partial E_s}{\partial \beta}\bigg>d\beta
 	\label{quasipre0}
 \end{equation}
 
 Thus the pressure gets an additional term due to temperature dependent quasiparticle energies. The first term is just an ideal gas pressure $P_\text{id}$. 
 
 To get $\mathcal{Q}_\beta$ we note that the entropy density (at fixed $(N,V)$) should depend only on the number of degress of freedom in the system. Hence, unlike pressure, it should not get any additional contribution due to temperature dependent quasiparticle energies. This condition is also maintained in our model.  For instance, at $\mu=0$, the entropy density of the system is 
 
 \begin{eqnarray}
s&=&\bigg(\frac{\partial P}{\partial T}\bigg)_{N,V}\\
&=&\frac{1}{V}\text{ln} \bigg[\sum_{r,s}\text{exp}(-\alpha N_r-\beta E_s) \bigg]+ \frac{1}{VT}\frac{\sum_{s,r}E_s e^{-\beta E_{r}-\alpha N_{s}}}{\mathcal{Z}}\label{entr2}\\
&=& \frac{P_{id}}{T}+\frac{\bar E}{VT}\label{entr3}\\
&=& \frac{P_{id}}{T}+\frac{E_{id}}{VT}\label{entr4}
 	\end{eqnarray}
 
 where $E_{id}$ is the idea gas energy. Adding and subtracting the term $\int \beta \bigg<\frac{\partial E_s}{\partial \beta}\bigg>d\beta$ on the right had side of  Eq.(\ref{entr3}) we get the entropy as
 \begin{equation}
 	S=\frac{PV}{T}+\frac{\bar E}{T}-\int \beta \bigg<\frac{\partial E_s}{\partial \beta}\bigg>d\beta
 	\end{equation}
 
 Comparing with  Eq. (\ref{1stlaw1}), we get desired expression for $\mathcal{Q}_{\beta}$ as:
 
 \begin{equation}
 	\mathcal{Q}_{\beta}=-\int \beta \bigg<\frac{\partial E_s}{\partial \beta}\bigg>d\beta
 	\label{Qbeta}
 	\end{equation}
 
Note that $d \mathcal{Q}_{\beta}$ term in the first law  ensures that the entropy of the system  remain independent of medium dependence of quasiparticle masses.

 
\section{ Thermodynamics of  gluon plasma}
\label{secIII}
In this section we apply the quasiparticle model formulated in Sec.\ref{secII} to a pure SU(3) Yang-Mills plasma.
$q-potential$ (\ref{qdef1}) can be written in terms of single particle energies as

\be
q=- \sum_{0}^{\infty} \text{ln}\bigg(1- z e^{-\beta\epsilon_k}\bigg)
\ee

where upper (lower) sign corresponds to bosons (fermions) and $z \equiv e^{\mu/T}$ is called fugacity. For gluons $\mu=0$, hence $z=1$. $\epsilon_k$ is the single particle energy of quasiparticles (gluons) with  temperature dependent mass:

\be
\epsilon_k(T)=\sqrt{k^2+m^2(T)}
\ee

where $k$ is the momentum and $m$ is the mass.




Pressure of a quasiparticle gas of gluons, in the thermodynamic limit, is written as

\begin{equation}
	P(T)=- \frac{g_f}{2\pi^2}\int_0^\infty dk \:k^2 \text{ln}\bigg[1-\text{exp}\bigg(-\frac{\epsilon_k(T)}{T}\bigg)\bigg]-\Phi(T)
\end{equation}

where,
\begin{equation}
\Phi(T)=	-\frac{g_f}{2\pi^2}\int_{0}^{T}dT^{'}  m\:\frac{dm}{dT^{'}} \int_{0}^{\infty} \frac{dk\: k^2}{\epsilon_k(k,T^{'})}\frac{1}{\text{exp}\bigg[\epsilon_k(k,T^{'})/T^{'}\bigg]-1}
\end{equation}

Energy density is given by,

\be
\varepsilon(T)=\frac{g_f}{2\pi^2}\int_0^\infty dk \:k^2 \frac{\epsilon_k(T)}{\text{exp}\bigg[\frac{\epsilon_k(T)}{T}\bigg]-1}
\ee

Specific quansiparticle model is characterized by different ways of parametrizing thermal quasiparticle mass. In this work we take\cite{Bannur:2012cf}

\be
m(T) = \frac{a}{(\frac{T}{T_c}-1)^c}+b(T/T_c)\frac{T}{T_c}
\ee
where $b(T/T_c)/T_c=\sqrt{2\pi\alpha_s(T/T_c)}$. Two-loop running coupling constant $\alpha_s(T)$ is given by

\be
\alpha_s(T)=\frac{6\pi}{(33-n_f)\text{ln}(T/\Lambda_T)}\bigg(1-\frac{3(153-19n_f)}{(33-2n_f)^2}\frac{\text{ln}(2\text{ln}(T/\Lambda_T))}{\text{ln}(T/\Lambda_T)}\bigg)
\ee

where $T_c/\Lambda_T=0.79$ from the LQCD results of reference \cite{Borsanyi:2012ve}. Further, for gluon plasma we may take $c=0.41$. Remaining parameter 'a' is fitted to reproduce LQCD results of reference \cite{Borsanyi:2012ve}.

\begin{figure}[h]
\vspace{-0.4cm}
\begin{center}
\begin{tabular}{c c}
\includegraphics[width=8cm,height=6cm]{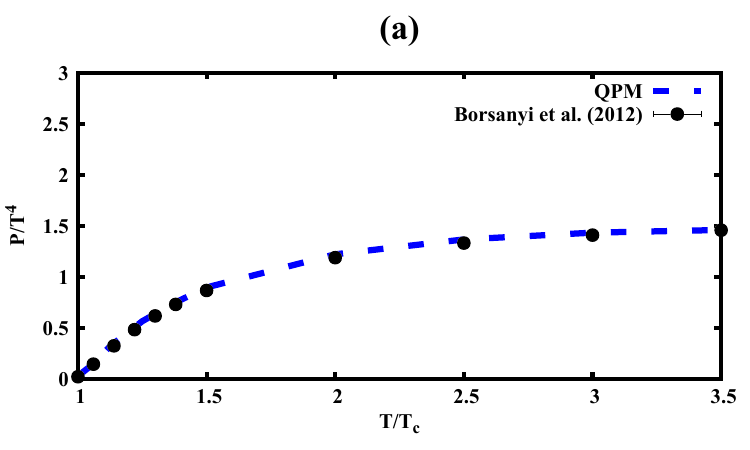}&
 \includegraphics[width=8cm,height=6cm]{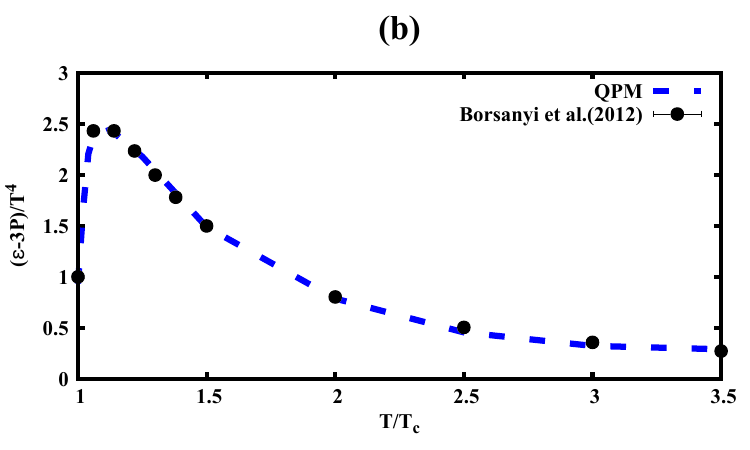}\\
  \end{tabular}
  \caption{(Color online)Left panel shows normalized pressure of SU(3) pure gluon plasma as a function of temperature (dashed blue curve). Right panel shows normalized interaction measure of SU(3) pure gluon plasma (dashed blue curve). Black dots corresponds to LQCD simulations of pure SU(3) gluodynamics\cite{Borsanyi:2012ve}.} 
\label{EoS_gluon}
  \end{center}
 \end{figure}
 
 Fig.\ref{EoS_gluon} shows scaled pressure and interaction measure as a function of temperature. We have compared our quasiparticle model (blue dashed curve) with the lattice QCD data of SU(3) gauge theory\cite{Borsanyi:2012ve}. Our model is in agreement with the lattice data over wide range of temperatures.

\section{Summary and conclusion}
\label{secIV}
In this paper we have given an alternative formulation of the quasiparticle model of quark gluon plasma with medium dependent dispersion relation.  We proposed a modification in the first law of thermodynamics  by including  temperature dependent term when single particle energies themselves are temperature dependent.  With this modified first law, we derived an expression for pressure starting from Kramer's  $q-potential$. We found that, to maintain thermodynamic consistency,  pressure receives an additional term solely due to medium dependent dispersion relation. The energy density, by definition, remains a sum over all the single particle energies with appropriate weight factor. Finally we compared our  the LQCD simulation results of SU(3) pure gluodynamics and  found excellent  agreement with it.

\bibliography{bib_qpm}
\end{document}